\documentclass[conference]{IEEEtran}
\usepackage{amsmath,subfigure}
\usepackage{graphicx}
\usepackage{grffile}
\usepackage{algpseudocode}
\usepackage{algorithm}
\usepackage{epstopdf}

\newtheorem{theorem}{Theorem}[section]

\newcommand{\qed}{\nobreak \ifvmode \relax \else
      \ifdim\lastskip<1.5em \hskip-\lastskip
      \hskip1.5em plus0em minus0.5em \fi \nobreak
      \vrule height0.75em width0.5em depth0.25em\fi}

\begin{document}
%
\title{Efficient Iterative Decoding of LDPC in the Presence of Strong Phase Noise}


\author{\IEEEauthorblockN{Shachar Shayovitz}
\IEEEauthorblockA{Department of Electrical Engineering\\
Tel Aviv University\\
Email: shachars@post.tau.ac.il}
\and
\IEEEauthorblockN{Dan Raphaeli}
\IEEEauthorblockA{Department of Electrical Engineering\\
Tel Aviv University\\
Email: danr@eng.tau.ac.il}}

\maketitle

\begin{abstract}
In this paper we propose a new efficient message passing algorithm for decoding LDPC transmitted over a channel with strong phase noise. The algorithm performs approximate bayesian inference on a factor graph representation of the channel and code joint posterior. The approximate inference is based on an improved canonical model for the messages of the Sum \& Product Algorithm, and a method for clustering the messages using the directional statistics framework. The proposed canonical model includes treatment for phase slips which can limit the performance of tracking algorithms. We show simulation results and complexity analysis for the proposed algorithm demonstrating its superiority over some of the current state of the art algorithms.

\end{abstract}
\section{Introduction}
\label{sec:intro}

Many satellite communication systems operating today employ low cost upconverters or downconverters which create phase noise. This noise can severely limit the information rate of the system and pose a serious challenge for the detection systems. Moreover, simple solutions for phase noise tracking such as PLL either require low phase noise or otherwise require many pilot symbols which reduce the effective data rate.

In the last decade we have witnessed a significant amount of research done on joint estimation and decoding of phase noise and coded information. These algorithms are based on the factor graph representation of the joint posterior distribution. This framework proposed in \cite{worthen2001}, allows the design of efficient message passing algorithms which incorporate both the code graph and the channel graph. The use of LDPC or Turbo decoders, as part of iterative message passing schemes, allows the receiver to operate in low SNR regions while requiring less pilot symbols.
The best known message passing algorithm for phase noise channels quantizes the phase noise and performs good approximation of the sum \& product algorithm (SPA). This algorithm (called DP - discrete phase in this paper) requires large computational resources to reach high accuracy, rendering it not practical for some real world applications.

In \cite{barb2005}, an algorithm which efficiently balances the tradeoff between accuracy and complexity was proposed (called BARB in this paper). BARB uses Tikhonov distribution parameterizations (canonical model) for all the SPA messages concerning a phase node. However, the approximation as defined in \cite{barb2005}, is only good when the information from the LDPC decoder is good (high reliability). In the first iteration the approximation is poor, and in fact is correct only for pilot symbols. The messages related to the received symbols which are not pilots are essentially zero (no information). This inability to accurately approximate the messages in the first iterations causes many errors and can create error floor. This problem is intensified when using either low code rate or high code rate. In the first case, it is since the pilots are less significant, since their energy is reduced. In the second case, the poor estimation of the symbols far away from the pilots cannot be overcome by the error correcting capacity of the code.
In order to overcome this limitation, BARB relies on the insertion of frequent pilots to the transmitted block causing a reduction of the information rate.

In order to improve the Tikhonov approximation, in this paper we suggest to avoid approximating the messages related to the received symbols, but to approximate only phase posterior messages by a Tikhonov distribution.
A major limitation of the resulting Tikhonov approximation is its sensitivity to phase slips. If the canonical model used to approximate the phase posterior messages is a Tikhonov distribution, then each message can only have one maximum point (called Phase Hypotheses). Therefore, all the hypotheses (mean values of the Tikhonov) across the code block describe a tracking plot of the phase noise (called Phase Trajectory). Phase slips occur when the estimated phase trajectory has an ambiguity, which can happen if different constellation symbols have similar likelihoods. Since the canonical model approximates this trajectory using a Tikhonov distribution, information is lost, and finally results with tracking a wrong trajectory, a phenomena that resembles phase slip in PLLs. We treat this problem by modifying the canonical model and adding a flat term i.e. uniform pdf to represent the possibility that the phase is not represented by the tracked trajectory. The scaling of this uniform pdf can be viewed as the probability of a phase slip.

In this paper, an iterative message passing algorithm is proposed, which uses a modified canonical model for the approximation of messages in the SPA. This algorithm needs fewer pilots and can operate in a wider range of code rates than BARB. The algorithm does not approximate the received symbols phase information, but applies a Tikhonov approximation later, on the forward and backward recursions wherein the phase estimation is updated. At each such recursion step, the posterior probability is a mixture of Tikhonov distributions. We apply a clustering algorithm based on the KL divergence in order to select which of the components of the mixture are going to be approximated by the Tikhonov and the rest will be left over as the flat hypothesis. The modified canonical model approach enables us to track a phase trajectory while maintaining a level of confidence for the tracked trajectory. This approach proves to be robust to phase slips and provides a high level of accuracy while keeping a low computational load.

An approximation of a mixture of Tikhonov pdfs to a single Tikhonov pdf optimal in sense of KL has not existed before, and required a new derivation, partially presented in this paper.

The resulting algorithm was shown in simulation to provide very good performance in high phase noise level and very close to the performance of the optimal algorithm even when very few pilots are present and the code rate is high.

\section{Preliminaries}

\subsection{Directional Statistics}
\label{sec:pagestyle}

Directional statistics is a branch of mathematics which studies random variables defined on circles and spheres. For example, the probability of the wind to blow at a certain direction. The circular mean and variance of a circular random variable $\theta$, are defined in \cite{mardia2000}:
\begin{equation}\label{tikh_mu}
    \mu_{Circular} = \arg[E(e^{j\theta})]
\end{equation}
\begin{equation}\label{tikh_var}
    \sigma^{2}_{Circular} = 1- E(cos(\theta-\mu_{Circular}))
\end{equation}

We define the operator $g(\theta) = \textsf{CMVM}[f(\theta)]$ (Circular Mean and Variance Matching), to take a circular pdf - $f(\theta)$ and create a Tikhonov pdf $g(\theta)$ with the same circular mean and variance.
Let $ g(\theta)$ be a Tikhonov distribution:
\begin{equation}\label{tikh_def}
   g(\theta) = \frac{e^{Re[k_{g}e^{-j(\theta-\mu_{g})}]}}{2\pi I_{0}(k_{g})}
\end{equation}

The following relations are obtained:
\begin{equation}\label{mu_circ}
   \mu_{g} = \mu_{Circular}(f)
\end{equation}
\begin{equation}\label{var_circ}
   \frac{I_{1}(k_{g})}{I_{0}(k_{g})} = 1 - \sigma^{2}_{Circular}(f)
\end{equation}

An alternative formulation for the Tikhonov pdf uses a single complex parameter $Z = ke^{j\mu}$

\subsection{Optimal Approximation of Tikhonov Mixture }
\label{ssec:subhead}
In this section we will show that the nearest Tikhonov distribution to a Tikhonov mixture (in a Kullback Liebler (KL) sense), has its circular mean and variance matched to those of the mixture.
The Kullback Liebler (KL) divergence \cite{KL1951} is a common information theoretic measure of similarity between probability distributions, which is defined as:
\begin{equation}\label{KL1}
    D(f||g) = \int_0^{2\pi}f(\theta)\log \frac{f(\theta)}{g(\theta)} d\theta
\end{equation}

\begin{theorem}
\emph{(CMVM):}
\label{mix_tikh_thr}
Let $f(\theta)=  \sum_{i =1}^{N}\alpha_{i}\frac{e^{Re[z_{i} e^{-j\theta}]}}{2\pi I_{0}(|z_{i}|)}$ be a Tikhonov mixture.
Then the Tikhonov distribution $g(\theta)$ which minimizes $D(f||g)$ is
\begin{equation}\label{CMVM_thr}
     g(\theta) = \textsf{CMVM}[f(\theta)]
\end{equation}
\end{theorem}

\begin{IEEEproof}

We will only provide an outline of the proof due to limited space.
Let $g(\theta) = \frac{e^{Re[ke^{-j(\theta-\mu)}]}}{2\pi I_{0}(k)}$

We wish to find
\begin{equation}\label{KL1}
    [\mu^{*},k^{*}] = arg\min_{\mu,k} D(f||g)
\end{equation}
The optimal $g(\theta)$ satisfies
\begin{equation}\label{diff0}
    \frac{\partial D(f||g)}{\partial k} = 0
\end{equation}

Skipping a few algebraic steps and using (\ref{tikh_var}), we get
\begin{equation}\label{eq11}
    \frac{I_{1}(k^{*})}{I_{0}(k^{*})} = \sum_{i =1}^{N}\alpha_{i}Re(e^{j(\mu^{*}-\mu_{i})})\frac{I_{1}(k_{i})}{I_{0}(k_{i})}
\end{equation}
where $\mu_{i} = \arg(z_{i})$ and $k_{i} = |z_{i}|$, parameters of the i'th Tikhonov distribution (mode) in the mixture $f(\theta)$. This is basically matching the circular variance of the mixture to the Tikhonov distribution.

The optimal $g(\theta)$ simultaneously needs to satisfy
\begin{equation}\label{mu_max}
    \mu^{*} = \arg \min_{\mu}{D(f||g)}
\end{equation}
Inserting $g(\theta)$ into (\ref{mu_max}) and skipping several steps we get
\begin{equation}\label{mean_max}
    {\mu^{*}} = \arg{{\sum_{i =1}^{N}\alpha_{i}\frac{I_{1}(k_{i})}{I_{0}(k_{i})}e^{j\mu_{i}}}}
\end{equation}
Which is basically a weighted vector sum of the expectations of each mode in the mixture. \qed
\end{IEEEproof}

\section{System Model and Previous Work}
\label{sec:format}

The system model and factor graph representation as in \cite{barb2005} are considered. For the sake of clarity, we will very briefly review the model and iterative algorithm.
We consider the transmission of a sequence of complex modulation symbols $\mathbf{c} = (c_{0},c_{1},...,c_{K-1})$
over an AWGN channel affected by carrier phase noise. We assume the symbols are drawn independency from an MPSK constellation. The discrete-time baseband complex equivalent channel model at the receiver is given by:
\begin{equation}\label{sys_model}
    r_{k} = c_{k}e^{j\theta_{k}}+n_{k} \;\;\;\;  k=0,1,...,K-1.
\end{equation}
The phase noise stochastic model is a wiener process
\begin{equation}\label{weiner}
    \theta_{k} = \theta_{k-1} + \Delta_{k}
\end{equation}
where ${\Delta_{k}}$ is a real, i.i.d gaussian sequence with $\Delta_{k} \sim \textsl{N}(0,\sigma_{\Delta}^{2})$.

The resulting Sum \& Product messages are computed by
\begin{equation}\label{pf}
    p_{f}(\theta_{k}) = \int_{0}^{2\pi}p_{f}(\theta_{k-1})p_{d}(\theta_{k-1})p_{\Delta}(\theta_{k}-\theta_{k-1})d\theta_{k-1}
\end{equation}
\begin{equation}\label{pb}
    p_{b}(\theta_{k}) = \int_{0}^{2\pi}p_{f}(\theta_{k+1})p_{d}(\theta_{k+1})p_{\Delta}(\theta_{k+1}-\theta_{k})d\theta_{k+1}
\end{equation}
\begin{equation}\label{pd}
    p_{d}(\theta_{k}) = \sum_{m=0}^{M-1} P_{d}(c_{k}=e^{j\frac{2\pi m}{M}}) f_{k}(c_{k},\theta_{k})
\end{equation}
\begin{equation}\label{Pu}
    P_{u}(c_{k}) = \int_{0}^{2\pi}p_{f}(\theta_{k})p_{b}(\theta_{k})f_{k}(c_{k},\theta_{k})d\theta_{k}
\end{equation}
where $M$,$r_{k}$ and $\sigma^{2}$ are the constellation order, received base band signal and the AWGN variance respectively. The messages $p_{d}$,$p_{b}$,$P_{d}$,$p_{f}$ and the functions $p_{\Delta}(\theta_{k})$, $f_{k}(c_{k},\theta_{k})$ are defined in \cite{barb2005}.

Due to the fact that the phase symbols are continuous random variables, a direct implementation of these equations is not possible and approximations are unavoidable. In \cite{barb2005}, a Tikhonov approximation is used for all the messages in the SPA which leads to a very simple and fast algorithm.

In the following sections, a new algorithm will be presented, which approximates the SPA messages using the directional statistics framework.

\section{Approximating the SPA messages}
\label{sec:pagestyle}

In \cite{barb2005}, the messages $p_{f}(\theta_{k})$, $p_{b}(\theta_{k})$ and $p_{d}(\theta_{k})$ were approximated by Tikhonov distributions and a message passing algorithm was derived based on the SPA recursion equations. In the first iteration, when there is no information from the LDPC decoder, this approximation provides poor results since $p_{d}(\theta_{k})$ is approximated as a uniform pdf. In order to improve accuracy, one can suggest a different approximation by realizing that there is no need to approximate $p_{d}(\theta_{k})$.

\begin{figure}
  \includegraphics[width=8.5cm]{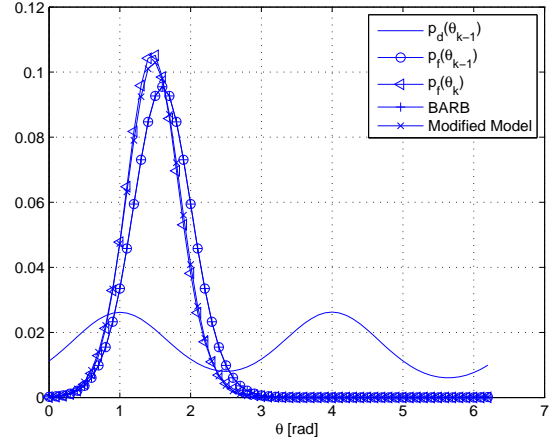}\\
  \caption{Comparison between approximation methods}\label{fig:approx}
\end{figure}

Decoding the LDPC code symbols only requires $P_{u}(c_{k})$ and the phase messages act behind the scenes. Therefore, in the computation of (\ref{pf}) and (\ref{pb}), $p_{d}(\theta_{k})$ can be used without approximation. Subsequently, only (\ref{pf}) and (\ref{pb}) need to be approximated as Tikhonov. This modification of BARB is shown in figure \ref{fig:approx} for very low phase noise variance. BARB estimates $p_{f}(\theta_{k})$ as $p_{f}(\theta_{k-1})$ because it has no data from $p_{d}(\theta_{k-1})$. Our modification approximates $p_{f}(\theta_{k})$ using the exact $p_{d}(\theta_{k-1})$. The results show that our modification is better.

However, This approximation creates a severe performance degradation due to phase slips. In this section we will present a different canonical model for the SPA messages which is robust to phase slips.

\subsection{Modified Canonical Models}
In this paper, a new approach, more robust to phase slips, is presented which utilizes a modified canonical model:
\begin{equation}\label{new_pf}
    p_{f}(\theta_{k-1}) = \tilde{\alpha}_{k-1}\tilde{p}_{f}(\theta_{k-1})+(1-\tilde{\alpha}_{k-1})\frac{1}{2\pi}
\end{equation}
\begin{equation}\label{new_pb}
    p_{b}(\theta_{k-1}) = \tilde{\beta}_{k-1}\tilde{p}_{b}(\theta_{k-1})+(1-\tilde{\beta}_{k-1})\frac{1}{2\pi}
\end{equation}
Where:
\begin{equation}\label{pf11}
    \tilde{p}_{f}(\theta_{k-1}) = \frac{e^{Re[Z^{f}_{k-1}e^{-j\theta_{k-1}}]}}{2\pi I_{0}(|Z^{f}_{k-1}|)}
\end{equation}
\begin{equation}\label{pb11}
    \tilde{p}_{b}(\theta_{k-1}) = \frac{e^{Re[Z^{b}_{k-1}e^{-j\theta_{k-1}}]}}{2\pi I_{0}(|Z^{b}_{k-1}|)}
\end{equation}

And $\tilde{\alpha}_{k-1}$, $\tilde{\beta}_{k-1}$ are real numbers between $0$ and $1$.

The modified canonical model can be viewed as a mixture of two distributions:
\begin{enumerate}
  \item Tikhonov: $\angle(Z^{f}_{k-1})$ is the forward phase noise hypotheses for $\theta_{k-1}$ and $|Z^{f}_{k-1}|$ is the confidence level of this hypotheses (inverse to the estimation variance).
  \item Uniform: representing all the other possible hypotheses not represented by the Tikhonov above.
\end{enumerate}

In this manner we are able to more accurately estimate the phase posterior and be robust to phase slip events.
We will show the derivations only for $p_{f}$, but the same applies for $p_{b}$ with a proper change in indexing.

In the previous section we have shown that the best way, in the KL sense, to approximate a Tikhonov mixture using a single Tikhonov is to match their circular mean and variance.

Inserting the modified canonical model (\ref{new_pf}) in (\ref{pf}), results in a a sum of two Tikhonov mixtures, both mixtures are in the order of the constellation size.
\begin{equation}\label{mix_pf_comp}
p_{f}(\theta_{k}) = \tilde{\alpha}_{k-1}M_{1}(\theta_{k}) + (1-\tilde{\alpha}_{k-1})M_{2}(\theta_{k})
\end{equation}
Where:
\begin{equation}\label{mix_pf1}
M_{1}(\theta_{k}) = \int_{0}^{2\pi}\tilde{p}_{f}(\theta_{k-1})p_{d}(\theta_{k-1})p_{\Delta}(\theta_{k}-\theta_{k-1})d\theta_{k-1}
\end{equation}
\begin{equation}\label{mix_pf2}
M_{2}(\theta_{k}) = \int_{0}^{2\pi}\frac{1}{2\pi}p_{d}(\theta_{k-1})p_{\Delta}(\theta_{k}-\theta_{k-1})d\theta_{k-1}
\end{equation}

The modified canonical model restricts (\ref{mix_pf_comp}) to have one Tikhonov distribution, thus only one phase trajectory can be tracked. Since the tracked phase trajectory was approximated using $\tilde{p}_{f}(\theta_{k-1})$, then the next step is to process only the mixture (\ref{mix_pf1}). It is important to understand that not all the components in (\ref{mix_pf1}) represent the same phase trajectory and there may be several hypotheses which split the tracked trajectory. Therefore, a selection algorithm must be employed which selects only one trajectory from (\ref{mix_pf1}) (which may be a clustering of several mixture components (hypotheses)). The selected components are clustered to a Tikhonov distribution while the rest of the components of (\ref{mix_pf_comp}) transfer their energy to the flat hypotheses.

\subsection{Component Selection and Clustering Algorithm}

In this section we propose an algorithm which selects components from the Tikhonov mixture (\ref{mix_pf_comp}) and clusters them to a single Tikhonov distribution. Note that $D(f||g)$ is the KL divergence between $f$ and $g$.
We define the input mixture to the algorithm as:
\begin{equation}\label{mix_pf3}
f(\theta)=  \sum_{i =1}^{N}a_{i}f_{i}(\theta)
\end{equation}

Where:
\begin{equation}\label{f_i}
f_{i}(\theta) =  \frac{e^{Re[z_{i} e^{-j\theta}]}}{2\pi I_{0}(|z_{i}|)}
\end{equation}

In order to combat phase slip events, the input (\ref{mix_pf3}) can be one of two mixtures: First, when the previous processed symbol was not a pilot, then mixture (\ref{mix_pf3}) is (\ref{mix_pf1}). Second, in case the previous symbol was a pilot then (\ref{mix_pf3}) is (\ref{mix_pf_comp}). It is assumed that if a phase slip has occurred, then the pilot will correct the tracked phase trajectory, so in the second case, the probability of the new hypotheses is 1: $\tilde{\alpha}_{k}=1$. This is a reasonable assumption in high SNR.

\begin{algorithm}
\caption{Component Selection \& Clustering Algorithm}
\label{select_algo}
\begin{algorithmic}
\State $lead \gets argmax_{i}\{\frac{a_{i}}{|z_{i}|^{-1}}\}$
\State $idx \gets lead$
 \For{$i = 1 \to N$}
     \If {$D(f_{lead}(\theta) || f_{i}(\theta)) \leq T_{D}$}
         \State $idx \gets [idx , i]$
    \EndIf
\EndFor
\State $\tilde{p}_{f}(\theta_{k}) \gets CMVM(a(idx),f(idx))$
\State $\tilde{\alpha}_{k} \gets \tilde{\alpha}_{k-1} \sum{a(idx)}$
\end{algorithmic}
\end{algorithm}

The algorithm finds the component with the highest amplitude to variance ratio (amp2var) and clusters all the other components "close" to it (KL sense and under a specified threshold). The ratio amp2var describes the ratio between the likelihood of the hypotheses and its estimation error. The assumption is that all the clustered components represent the same phase trajectory. The likelihood $\tilde{\alpha}_{k}$ is computed using the weights of all the clustered mixture components. Therefore, the modified canonical model can be viewed as a weighted sum of mixture components which better estimates the phase posterior than a single Tikhonov distribution.

There is a tradeoff in the selection of the threshold $T_{D}$. It should be low enough so that two components associated with different phase trajectories are not clustered but high enough so that all the components representing the same phase trajectory will be clustered together.

\subsection{Using \textsf{CMVM} to Cluster Mixture Components}
Algorithm 1. selects a set $J$ of components indices from mixture (\ref{mix_pf1}). These components, represent the same phase estimate and are approximated using a single Tikhonov.
The mixture composed of components from (\ref{mix_pf1}) with indices from the set $J$ is:
\begin{equation}\label{mix_tikh_11}
    p^{J}_{f}(\theta_{k})=\sum_{l \in J}^{|J|}\alpha_{l}\frac{e^{Re[(\hat{Z}_{l})e^{-j\theta_{k}}]}}{2\pi I_{0}(|\hat{Z}_{l}|)}
\end{equation}
Where:
\begin{equation}\label{Z_fin}
    \hat{Z}_{l} = \frac{Z_{l}}{1+\sigma_{\Delta}^{2}|Z_{l}|}
\end{equation}

And:
\begin{equation}\label{Z}
    Z_{l} = Z^{f}_{k-1}+\frac{r_{k-1}e^{-j\frac{2\pi J(l)}{M}}}{\sigma^2}
\end{equation}
\begin{equation}\label{coeff}
    \alpha_{l} = \frac{1}{A}P_{d}(c_{k-1}=e^{j\frac{2\pi J(l)}{M}})\frac{I_{0}(|Z_{l}|)}{I_{0}(|Z^{f}_{k-1}|)}
\end{equation}
Where $A$ is a normalizing constant.

Using Theorem (\ref{mix_tikh_thr}) and skipping the algebraic details, the \textsf{CMVM} operator for (\ref{mix_tikh_11}), is:
\begin{equation}\label{cmvm_res}
    \textsf{CMVM}(p^{J}_{f}(\theta_{k})) = \frac{e^{Re[Z^{f}_{k}e^{-j\theta_{k}}]}}{2\pi I_{0}(|Z^{f}_{k}|)}
\end{equation}
Where:
\begin{equation}\label{Zf}
    Z_{k}^{f} = \hat{k}e^{j\hat{\mu}}
\end{equation}
And:
\begin{equation}\label{tikh_modify_eqs_mu}
    \hat{\mu} = \arg{{\sum_{l \in J}^{|J|}\alpha_{l}\frac{I_{1}(|Z_{l}|)}{I_{0}(|Z_{l}|)}e^{j\arg(Z_{l})}}}
\end{equation}
\begin{equation}\label{tikh_modify_eqs_k}
    \frac{1}{2 \hat{k}} = 1 - \sum_{l \in J}^{|J|}\alpha_{l}\frac{I_{1}(|Z_{l}|)}{I_{0}(|Z_{l}|)}Re[e^{j(\hat{\mu}-arg(Z_{l}))}]
\end{equation}

Similar results apply also for the backward recursion. The computational complexity of implementing a modified bessel function in a system is prohibitively expensive. We therefore present the following approximation
\begin{equation}\label{assumps3}
    \log(I_{0}(k)) \approx k-\frac{1}{2}\log(k)-\frac{1}{2}\log(2\pi)
\end{equation}
which holds for $k>2$, i.e. reasonably narrow distributions.

\begin{equation}\label{assumps1}
    I_{1}(x) = \frac{dI_{0}(x)}{dx}
\end{equation}
We find that
\begin{equation}\label{fracI1I0}
    \frac{I_{1}(k)}{I_{0}(k)} = \frac{d}{dk}(\log(I_{0}(k)))
\end{equation}
Therefore
\begin{equation}\label{eq5}
    \frac{I_{1}(k)}{I_{0}(k)} \approx 1-\frac{1}{2k}
\end{equation}

Thus, the simplified approximated versions of (\ref{tikh_modify_eqs_k}) and (\ref{tikh_modify_eqs_mu}) are
\begin{equation}\label{tikh_modify_eqs_mu_simp}
    \hat{\mu} = \arg[{{\sum_{l \in J}^{|J|}\alpha_{l}(1-\frac{1}{2|Z_{l}|})e^{j\arg(Z_{l})}}}]
\end{equation}
\begin{equation}\label{tikh_modify_eqs_k_simp}
    \frac{1}{2 \hat{k}} = 1 - \sum_{l \in J}^{|J|}\alpha_{l}(1-\frac{1}{2|Z_{l}|})\cos(\hat{\mu}-\arg(Z_{l}))
\end{equation}
We also use the approximation for the modified bessel function in the computation of $\alpha_{l}$.

\subsection{Computation of $P_{u}(c_{k})$}

Another advantage of using the modified canonical model comes to effect when we use the forward-backward scheduling of the message passing algorithm to compute the message $P_{u}(c_{k})$. This message describes the LLR of a code symbol and the correct estimation of this measure is crucial for the decoding of the LDPC.

We insert our modified canonical model (\ref{new_pf}), (\ref{new_pb}) into (\ref{Pu}):
\begin{equation}\label{Pu_new}
    P_{u}(c_{k}) = A + B + C + D
\end{equation}

Where:
\begin{equation}\label{A}
    A = \tilde{\alpha}_{k}\tilde{\beta}_{k}\int_{0}^{2\pi}(\tilde{p}_{f})(\tilde{p}_{b})f_{k}(c_{k},\theta_{k})d\theta_{k}
\end{equation}

\begin{equation}\label{B}
    B = \tilde{\alpha}_{k}(1-\tilde{\beta}_{k})\int_{0}^{2\pi}(\tilde{p}_{f})(\frac{1}{2\pi})f_{k}(c_{k},\theta_{k})d\theta_{k}
\end{equation}

\begin{equation}\label{C}
    C = (1-\tilde{\alpha}_{k})\tilde{\beta}_{k}\int_{0}^{2\pi}(\frac{1}{2\pi})(\tilde{p}_{b})f_{k}(c_{k},\theta_{k})d\theta_{k}
\end{equation}

\begin{equation}\label{D}
    D = (1-\tilde{\alpha}_{k})(1-\tilde{\beta}_{k})\int_{0}^{2\pi}(\frac{1}{2\pi})(\frac{1}{2\pi})f_{k}(c_{k},\theta_{k})d\theta_{k}
\end{equation}
Thus $P_{u}(c_{k})$ is a weighted summation of four components which can be interpreted as conditioning on the probability that a phase slip has occurred for each recursion (forward and backward). This will ensure that the computation of $P_{u}(c_{k})$ is based on the most reliable phase posterior estimations, even if a phase slip has occurred in a single recursion (forward or backward).

\section{Complexity}
As stated in the introduction, DP suffers from a complexity limitation. The DP algorithm quantizes the phase symbols and performs the SPA on high resolution values of the phase. In table 1. we compare the computational load for DP, BARB and the algorithm proposed in this paper.

\begin{table}[h]
\caption{Computational load per code symbol per iteration for M-PSK constellation}  

\centering  
\begin{tabular}{cccc}  
\\[1ex] \hline\hline                       
  &DP & BARB & Our Algorithm
\\ [0.5ex]
\hline\hline             
Operations & 13ML+10QL-9L-3M& 17M+11& 40M\\[1ex]
LUT & 3ML+2QL-3L-M& 3M+3 & 5M\\[1ex]
\hline                          
\end{tabular}
\label{tab:PPer}
\end{table}

$M$ is the constellation order, $L$ is the number of quantization levels and $Q$ is a parameter for the DP algorithm explained in \cite{barb2005}.

We can see that we can achieve a high level of accuracy while maintaining a low computational load. Therefore, the algorithm proposed in this paper provides an improved tradeoff between accuracy and complexity than BARB.

\section{Numerical Results}
\begin{figure}
  \includegraphics[width=8.5cm]{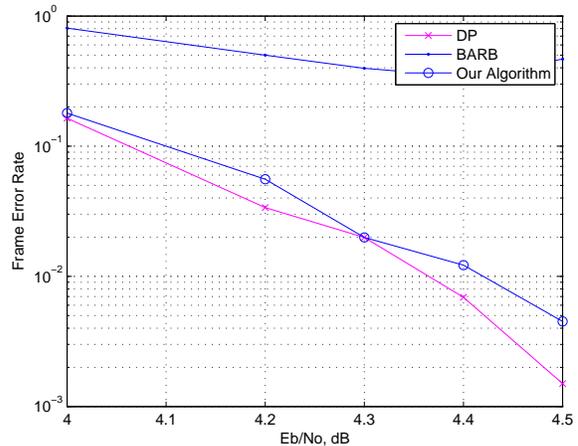}\\
  \caption{Frame error rate - BPSK and wiener phase noise}\label{fig:res}
\end{figure}

Monte Carlo simulation results for the various algorithms are shown  in Fig \ref{fig:res}. A length 4608 LDPC code with rate 0.889 was used. The bits were mapped to a BPSK constellation and the phase noise model used was a wiener process with $\sigma_{\Delta}=0.1$[rads/symbol]. A single pilot was inserted every 80 symbols and the parameter $T_{D}$ for the selection algorithm was chosen to be $2.2$. The DP algorithm was simulated using 8 quantization levels.

As shown in Fig \ref{fig:res}, BARB demonstrates a high error floor. This is because of the large phase noise variance and large spacing between pilots which causes the posterior estimation to become uniform and thus does not provide information for the LDPC decoder. The high code rate amplifies this problem. The new algorithm has a negligible with respect to DP algorithm.


\bibliographystyle{plain}
\bibliography{strings}

\begin{thebibliography}{1}

\bibitem{worthen2001}
W.~Stark A.~Worthen.
\newblock Unified design of iterative receivers using factor graphs.
\newblock {\em IEEE Transactions on Information Theory}, 47:843 --849, February
  2001.

\bibitem{barb2005}
A.~Barbieri G.~Colavolpe and G.~Caire.
\newblock Algorithms for iterative decoding in the presence of strong phase
  noise.
\newblock {\em IEEE Journal on Selected Areas in Communications}, 23:1748
  --1757, September 2005.

\bibitem{KL1951}
S.~Kullback and R.~A. Leibler.
\newblock On information and sufficiency.
\newblock {\em The Annals of Mathematical Statistics}, 22:79--86, March 1951.

\bibitem{mardia2000}
KV. Mardia and Jupp P.
\newblock {\em Directional Statistics}.
\newblock John Wiley and Sons Ltd., 2000.

\end{thebibliography}

\end{document}